\documentclass[prd,twocolumn,showpacs,superscriptaddress,preprintnumbers,nofootinbib,
amsmath,amssymb]{revtex4-1}

\usepackage{graphicx} % Required for inserting images
\usepackage{tikz}
\usetikzlibrary{arrows}

\usepackage[colorlinks=true,urlcolor=blue,linkcolor=blue,citecolor=blue]{hyperref}
\usepackage[normalem]{ulem}
\usepackage[capitalize]{cleveref}
 
\usepackage{comment}
\usepackage{flushend}

\usepackage[T1]{fontenc} % if needed

\usepackage{mathtools}
\usepackage{xcolor}
\usepackage{graphicx}% Include figure files
\usepackage{dcolumn}% Align table columns on decimal point
\usepackage{bm}% bold math
\usepackage{multirow}
\usepackage{physics}
\usepackage{color, soul}
\usepackage{epstopdf}
\usepackage{float}
\usepackage{subcaption} 
\newcommand{\Beq}{\begin{equation}\begin{aligned}}
\newcommand{\Eeq}{\end{aligned}\end{equation}}

\begin{document}

%\preprint{XXX}
\title{Small-Scale Clustering of Primordial Black Holes: The Little Red Dot Mass Function and the High-Redshift Galaxy Tension}
%\title{Small-Scale Clustering of Primordial Black Holes:\\ Mass Function and the Origin of JWST Little Red Dots} 

\author{Borui Zhang}
%\email{zhangbr22@mails.tsinghua.edu.cn}
\affiliation{Department of Physics, Tsinghua University, Beijing 100084, China}

\author{Wei-Xiang Feng}
\email{wxfeng@mail.tsinghua.edu.cn}
\affiliation{Department of Physics, Tsinghua University, Beijing 100084, China}

\author{Haipeng An}
\email{anhp@mail.tsinghua.edu.cn}
\affiliation{Department of Physics, Tsinghua University, Beijing 100084, China}
\affiliation{Center for High Energy Physics, Tsinghua University, Beijing 100084, China}

%\date{\today}

\begin{abstract}
Supermassive black holes (SMBHs) in ``little red dots'' (LRDs) discovered by the James Webb Space Telescope (JWST) may result from runaway mergers of primordial black holes (PBHs) in clusters---through long--short mode coupling on small scales in the early Universe. In this framework, we derive the SMBH mass function, together with the compactness and overmassive features of LRDs. We also estimate that the dense gas residing in PBH clusters is consistent with LRD observations. In addition, SMBHs formed from PBH clusters can help accelerate galaxy formation at high redshifts, thus alleviating tension with $\Lambda$CDM cosmology.
\end{abstract}

\maketitle

\section{Introduction}
Recent James Webb Space Telescope (JWST) observations have revealed a previously unseen population of compact (\(\lesssim\!100{\rm\,pc}\)), red, low-to-intermediate-luminosity active galactic nuclei (AGN), the so-called ``little red dots'' (LRDs)\;\cite{Matthee:2023utn,Pacucci:2023oci,Gentile:2024uiw,akins2024cosmos,williams2024galaxies}—identified over a redshift range \(z\sim4\textup{--}11\). The inferred masses of central supermassive black holes (SMBHs) in LRDs are typically \(M_\bullet\sim10^{5\textup{--}8}{\rm\,M}_\odot\)\;\cite{Harikane:2023aa,Maiolino:2023bpi,kocevski2023hidden,kokorev_uncover_2023,killi2024deciphering,kokorev_census_2024,wang_rubies_2024,Durodola:2024bom,Ananna:2024jug,kocevski2024rise,Furtak_2024,Maiolino:2025tih}, and they are about 1–2 orders of magnitude more abundant than quasars at similar redshifts. A substantial fraction of SMBHs in LRDs are ``overmassive'' relative to their measured stellar components, with mass ratios \(M_\bullet/M_\star\sim 0.01\textup{--}1\), about \(1\textup{--}2\) orders of magnitude compared to the local relations~\cite{kocevski2023rise,kocevski2024rise,harikane2023jwst,maiolino2024jades,Maiolino:2025tih}. Recent studies identifying black holes in LRDs as nearly naked further corroborate this overmassive feature~\cite{Maiolino:2025tih}. 

Since LRDs are likely the sites where the first black holes formed, their bizarre properties may provide the key to understanding the formation of massive black holes in the early Universe and their coevolution with the first-generation galaxies. In particular, SMBHs in LRDs could be of primordial origin~\cite{Zhang:2025tgm,Zhang:2025oyl,DeLuca:2025nao,Garcia-Bellido:2026wig,Zhang:2026fhn,Magaraggia:2026ngp}, known as primordial black holes (PBHs) resulting from high-density perturbations in the early Universe~\cite{Hawking:1974rv,Chapline:1975ojl,Sasaki:2018dmp,Carr:2020gox}. As they can span a wide range of masses during the radiation era, some of them can persist to late times and grow through accretion and mergers.

In Ref.\,\cite{Zhang:2025tgm}, we investigated the origin of SMBHs in LRDs within the PBH framework, aiming to evade the stringent CMB constraints~\cite{Sasaki:2018dmp,Carr:2020gox}. We proposed a scenario in which SMBHs form through runaway mergers of dense PBH clusters. Such clusters can arise from the modulation of the overdensity field at the PBH formation scale (\(k_s\)) by long-wavelength modes (\(k_l\)) on larger scales. This modulation can be generated through a variety of mechanisms, including local-type non-Gaussianity~\cite{Byrnes:2011ri,Byrnes:2012yx,Young:2013oia,Young:2014oea,Young:2015kda,Tada:2015noa,Franciolini:2018vbk,Desjacques:2018wuu,Ali-Haimoud:2018dau,Young:2019gfc,Suyama:2019cst,Atal:2020igj,DeLuca:2022bjs,DeLuca:2022uvz}, closed domain walls~\cite{Khlopov:2004sc,Dokuchaev:2004kr,Belotsky:2018wph}, multistream inflation~\cite{Ding:2019tjk,Huang:2023mwy}, long-range scalar forces~\cite{Amendola:2017xhl,Savastano:2019zpr,Flores:2020drq}, quantum diffusion~\cite{Ezquiaga:2019ftu,Ezquiaga:2022qpw,Animali:2024jiz}, or correlated bubble collisions~\cite{DeLuca:2021mlh}. In this framework, the formation of SMBHs within the first billion years can be explained, thereby accounting for the origin of SMBHs in LRDs. Moreover, runaway mergers of dense PBH clusters are expected to generate distinctive gravitational-wave signatures~\cite{Zhang:2025tgm}, which could distinguish them from other formation channels.

In this paper, we extend our study of the PBH-cluster scenario, aiming to explain the characteristics of LRDs and the JWST high-redshift galaxies. The SMBH mass function derived in our model is consistent with LRD observations when accretion effects are included. In particular, our model links the central black-hole mass to the properties of host halo, stellar mass, and gas-disk radius, due to the seed effect induced by the PBH clusters. This automatically explains the overmassive feature of SMBHs in LRDs. If LRDs are associated with the low-angular-momentum components of the seed-effect-induced haloes, then the observed compactness can also be explained. In addition, PBH clusters also generate seeds at the high-mass end, different from LRDs, which can accelerate early galaxy formation and reconcile the cumulative stellar mass density inferred from JWST.

The rest of the paper is organized as follows. In Sec.\,\ref{Sec:mass_function}, we derive the SMBH mass function from our model of PBH clusters. In Sec.\,\ref{Sec:seed_effect}, we explain the properties of LRD through the prescription of the seed effect. In Sec.\,\ref{Sec:LRD_dense_gas}, we show that our modeling may result in a dense gas density consistent with that in LRDs. Sec.\,\ref{Sec:high-z_galaxy} demonstrates that PBH clusters on the high-mass tail can facilitate the early galaxy formation observed by JWST. Sec.\,\ref{Sec:discussion} discusses the LRD abundance in our modeling. We conclude in Sec.\,\ref{Sec:conclusion}. Throughout the paper, we adopt $H_0=68{\rm\,km/s/Mpc}$, $\Omega_{\rm M}=0.31$, $\Omega_\Lambda=0.69$ at present from Planck cosmology~\cite{Planck:2018vyg}.

\section{Mass function of SMBHs from PBH clusters} 
\label{Sec:mass_function}
In this section, we derive the SMBH mass function in the PBH cluster model~\cite{Zhang:2025tgm}. We first derive the threshold for cluster formation and the properties of a cluster, including the cluster mass, the PBH number density, and the runaway merger timescale. We then establish the relation between the SMBH mass function and the probability distribution function of the cluster mass. By taking the effect of accretion into account, we obtain the SMBH mass function predicted in our model.

\subsection{The PBH cluster model} We assume that PBH clusters arise from long--short mode coupling. Models of long-mode modulation can be parametrized as \(\nu(\mathbf{x})=\nu\!\left(\Phi_l(\mathbf{x})\right)\), where \(\Phi_l(\mathbf{x})\) denotes the long-wavelength field assumed to follow Gaussian distribution and \(\nu \equiv \delta_c/\bar{\sigma}\) is the reduced threshold for PBH formation, with \(\delta_c\) the critical density fluctuation during the radiation era and \(\bar{\sigma}^2\) the corresponding variance averaged over the whole space. The initial local abundance of PBHs defined as the PBH energy fraction of the total radiation energy density is modulated by the long mode, given by~\cite{Atal:2020igj}
\begin{align}
\beta(\nu(\Phi_l(\mathbf{x})))=\frac{1}{2}\mathrm{erfc}\left(\frac{\nu(\Phi_l(\mathbf{x}))}{\sqrt{2}}\right)\,,
\end{align}
which manifests the initial clustering of PBHs, and one can obtain the cluster number density, mass and virial velocity from this quantity~\cite{Zhang:2025tgm}.

At the PBH formation time $a_{\rm form}$ (\(a\) is the scale factor), the local PBH energy density is 
\begin{align}
\rho_{\rm pbh}(a_{\rm form})=\beta(\nu(\Phi_l(\mathbf{x}))\rho_{\rm rad} (a_{\rm form})\;.
\end{align}
To characterize the formation threshold of PBH clusters as gravitational bound objects, we define the PBH cluster density contrast as
\begin{align}
\delta_{\rm cl}(a)\equiv \frac{\rho_{\rm pbh}(a)}{\rho_{\rm rad}(a)}\;.
\end{align}
A cluster can form only if it decouples from the Hubble flow, which requires two conditions to be satisfied. First, the PBH local energy density at the decoupling time $a_{\rm dec}$ must become comparable to the background radiation energy density, i.e., \( \rho_{\rm pbh}\simeq\rho_{\rm rad}\), which implies the threshold \(\delta_{\rm cl}(a_{\rm dec})\simeq1\)\footnote{This threshold, unlike the threshold of forming dark halos from the spherical collapse model, is independent of the cluster shape.}; and thus the decoupling scale factor \(a_{\rm dec}={a_{\rm form}}/{\beta(\nu(\Phi_l(\mathbf{x})))}\). Besides, the decoupling time should be earlier than the matter-radiation equality \(a_{\rm dec}<a_{\rm eq}\). This sets a lower bound on the initial local abundance $\beta$ and thus a threshold for the initial clustering \(\delta_{{\rm cl},i}\equiv\delta_{{\rm cl}}(a_{\rm form})=\beta(\Phi_l(\mathbf{x}))>a_{\rm form}/a_{\rm eq}\). Together with the relation
\({a_{\rm eq}}/{a_{\rm form}}=\left({3\times 10^{17} \mathrm{M}_\odot}/{m_{\rm pbh}}\right)^{{1}/{2}}\)~\cite{Ozsoy:2023ryl}, the lower bound becomes 
\begin{align}\label{Eq:lower_bound_cluster}
\delta_{\mathrm{cl},i}=\beta>\left(\frac{3\times 10^{17} \mathrm{M}_\odot}{m_{\rm pbh}}\right)^{-1/2}.
\end{align}
For PBHs of \(m_{\rm pbh}=30\,\mathrm{M}_\odot\) adopted in this work, the lower bound is \(\beta>10^{-8}\). To generate dense PBH clusters, \(\beta\) should be a monotonically decreasing function of \(\nu(\Phi_l)\). Therefore, the above inequality sets an upper bound on \(\nu(\Phi_l)\). Once the clusters form, we identify \(\rho_{\rm pbh}=\rho_{\rm cl}\) as the mass density of PBHs in a cluster.

To obtain the physical properties of the resulting cluster, for simplicity, we assume that all clusters are spherical and non-overlapping with a comoving scale approximately equal to that of the long-wavelength mode, i.e., 
\begin{align}
r_{\rm cl}\simeq \frac{1}{k_l}
\end{align}
with \(k_l\) the comoving momentum of the long mode. This assumption is reasonable for the long-wavelength perturbation taking an extreme (tail) value to form a cluster compact enough for runaway mergers~\cite{Zhang:2025tgm,Zhang:2026vjk}. Although this is not as extreme as that required to form a PBH,\footnote{For PBH formation, the abundance scales as \(\propto \mathrm{erfc}(\nu/\sqrt{2})\sim 10^{-17}\) for a characteristic variance of the long mode. Therefore, the PBH abundance from direct collapse of the long mode itself is essentially negligible.} it is still much larger than the variance. Therefore, the peak is expected to be approximately spherical according to Gaussian peak theory (see Ref.\,\cite{Bardeen:1985tr}), and the cluster overlapping can be neglected. Furthermore, we approximate the long-wavelength mode as (nearly) constant across its coherence scale, as expected for a Gaussian random field~\cite{Bardeen:1985tr}. Under these assumptions, we obtain the mass of a cluster as a function of the field value \(M_{\rm cl}=\left(4\pi/3\right)r_{\rm cl}^3 a_{\rm form}^3\beta(\nu(\Phi_l))\rho_{\rm rad}(a_{\rm form})\), and it can then be written as
\begin{align}\label{Eq:MCl(Phi_l)}
M_{\rm cl}
&\simeq 6.1\times10^{6}{\rm\,M}_\odot\left(\frac{r_{\rm cl}}{10^4\,\mathrm{pc}}\right)^3\left(\frac{m_{\rm pbh}}{30\, \mathrm{M}_\odot}\right)^{-1/2}\notag\\
&\times
\frac{\beta(\nu(\Phi_l))}{3.7\times10^{-7}}\;.
\end{align}
The number density of PBHs in a cluster after decoupling and virialization can be calculated as \(n_{\rm cl}=C{\rho_{\rm cl}(a_{\rm dec})}/{m_{\rm pbh}}\), or equivalently,
\begin{align}
n_{\rm cl}
&=10^8\,\mathrm{pc}^{-3}\left(\frac{C}{20}\right)\left(\frac{m_{\rm pbh}}{30\;\mathrm{M}_\odot}\right)^{3}\left(\frac{\beta(\nu(\Phi_l))}{3.7\times10^{-7}}\right)^4\,,
\end{align}
where the compactness parameter $C$ is adopted to characterize the cluster's overdensity after virialization~\cite{Kolb:1994fi}, and we have used the relation \({a_{\rm dec}}/{a_{\rm form}}={1}/{\beta}\) and \({a_{\rm eq}}/{a_{\rm form}}=\left({3\times10^{17}\,\mathrm{M}_\odot}/{m_{\rm pbh}}\right)^{{1}/{2}}\). Note that the number density $n_{\rm cl}\propto\beta^4$ thus is more sensitive to the value of the long-mode field compared to $M_{\rm cl}\propto\beta$. 

The runaway timescale of PBH mergers in a cluster is given by \(t_{\rm ra}=\tilde{t}_{\rm ra}t_{\rm upp}\) with \(t_{\rm upp}=1/(n_{\rm cl}\mathcal{K}_{00})\) being the upper bound~\cite{Zhang:2026vjk}. Here \(\mathcal{K}_{00}=2^{{9}/{14}}\mathcal{A}{(G^2/c^3)\,m_{\rm pbh}^2}\left({v_0}/{c}\right)^{-{11}/{7}}\) is the averaged merger cross-section of PBHs with \(\mathcal{A}=85^{2/7} (2\pi)^{11/14}\sqrt{3}\,\Gamma \left(5/7\right)\) and $v_0$ the PBH velocity dispersion in the cluster. The dimensionless runaway timescale \(\tilde{t}_{\rm ra}\simeq \mathcal{O}(10^{-2}\textup{--}10^{-1})\) can be obtained from our Monte Carlo simulations based on the Smoluchowski coagulation equation~\cite{Zhang:2026vjk}. Specifically, as the velocity dispersion of PBHs in the cluster is defined by \(v_0=\sqrt{{3}/{5}}\,v_{\rm vir}=\sqrt{{3}/{5}}\sqrt{{GM_{\rm cl}}/{r_{\rm vir}}}\), with $v_{\rm vir}$ the virial velocity within the virial radius  $r_{\rm vir}=\left[{3M_{\rm cl}}/({4\pi m_{\rm pbh}n_{\rm cl}})\right]^{-{1}/{3}}$, the runaway timescale can be written in terms of the cluster mass and number density, so that
\(t_{\rm ra}\propto n_{\rm cl}^{-1}v_{\rm vir}^{11/7}\propto M_{\rm cl}^{11/21}n_{\rm cl}^{-31/42}\). After substituting the explicit expressions of \(M_{\rm cl}\) and \(n_{\rm cl}\), we obtain
\begin{align}
t_{\rm ra}&\simeq 1.1{\rm\,Gyr}\left(\frac{\tilde{t}_{\rm ra}}{0.1}\right)\left(\frac{m_{\rm pbh}}{30\;\mathrm{M}_\odot}\right)^{3/14}\left(\frac{r_{\rm cl}}{10^4{\rm\,pc}}\right)^{11/7}\notag\\
&\quad \times\left(\frac{\beta(\nu(\Phi_l))}{3.7\times10^{-7}}\right)^{-17/7},
\end{align}
which allows us to determine if PBH clusters can result in central SMBHs.

\subsection{The mass function of SMBHs}
Since the mass function of SMBHs is associated with the mass distribution function of PBH clusters, we need to first calculate the probability distribution function (PDF) of the cluster mass. Given the PDF of $\Phi_l$, the PDF of the cluster mass can be obtained as\,\footnote{If $y=f(x)$ is given as a function of $x$, and the inverse function $x=f^{-1}(y)$ is associated with the PDF of $x$ as \(P_{X}(x)\), the PDF of $y$ is given by \(P_{Y}(y)=P_{X}(f^{-1}(y))\big|{{\rm d}f^{-1}(y)}/{\rm d}y\big|\).}
\begin{align}
P(M_{\rm cl})=\frac{1}{\sqrt{2\pi}\sigma_l}\exp\left[-\frac{\Phi_l^2(M_{\rm cl})}{2\sigma_l^2}\right]\bigg|\frac{{\rm d}\Phi_l(M_{\rm cl})}{{\rm d}M_{\rm cl}}\bigg|\,,
\end{align}
where $\Phi_l$ is expressed as a function of cluster mass and \(\sigma_l^2\) is the corresponding variance.

As the observed SMBH mass function is defined within a specific redshift interval, we consider only those clusters that undergo runaway mergers into SMBHs in this interval. Alternatively, one may impose a bound on the progenitor mass by requiring the runaway timescale to be shorter than the cosmic time at the redshift of interest. Consequently, $\beta(\nu(\Phi_l))$ further acquires a lower bound determined by the redshift at which the SMBH appears.

We require that the cluster undergo runaway mergers and collapse into an SMBH before it is observed by JWST, that is \( t_{\rm ra}(\nu(\Phi_l)) < t(z_{\rm obs})\). Thus, the lower bound of \(\beta\) becomes 
\begin{align}\label{Eq:lower_bound_beta}
&\beta\gtrsim3.7\times10^{-7}\notag\\
&\times\left[\frac{1.1{\rm\,Gyr}}{t(z_{\rm obs})}\left(\frac{\tilde{t}_{\rm ra}}{0.1}\right)\left(\frac{m_{\rm pbh}}{30\,\mathrm{M}_\odot}\right)^{3/14}\!\left(\frac{r_{\rm cl}}{10^4\,{\rm pc}}\right)^{11/7}\right]^{7/17}.
\end{align}
Since most LRDs are observed at redshifts \(z\gtrsim 4\), we adopt a cutoff redshift of \(z_{\rm obs}\simeq 4\). For typical values \(m_{\rm pbh}=30{\rm\,M}_\odot\), \(r_{\rm cl}=10^{4}\,\mathrm{pc}\), and \(\tilde{t}_{\rm ra}=0.1\), we find that the lower bound of \(\beta\) becomes \(\beta\gtrsim 1.2\times10^{-7}\), which is greater than the bound in Eq.\,\ref{Eq:lower_bound_cluster}. Since the physically relevant lower bound is the maximum of these two constraints, we adopt the value derived from the observed redshift range as our lower bound on \(\beta\).

For a concrete model, we consider a simple linear parametrization of the long-mode spatial modulation, given by\;\cite{Atal:2020igj,Zhang:2025tgm}: 
\begin{align}\label{Eq:long_short_coupling}
\nu(\mathbf{x})=\nu_{g}\left(1+\eta\,\Phi_{l}(\mathbf{x})\right)\,,
\end{align}
where \(\eta\) is the coupling strength between long and short modes, \(\Phi_l(\mathbf{x})\) is the long-mode field, and \(\nu_g \equiv \delta_c / \bar{\sigma}_{s}\) is the reduced threshold of the short mode without modulation, with \(\delta_c \simeq 0.414\) during the radiation era\;\cite{Harada:2013epa}, and \(\bar{\sigma}_{s}^2 = \langle\delta_s^2\rangle\) the variance of \(\delta_s\) averaged over the whole space. Both the long- and short-mode density fields are assumed to follow Gaussian distributions. In this study, we follow the parameter set in our recent work~\cite{Zhang:2025tgm}: fix $\nu_g=8.5$ to generate a sufficiently large initial clustering amplitude and to achieve an adequate PBH abundance, resulting in $\bar{\sigma}_s\simeq0.0487$. The variance of the long-mode, $\sigma_l^2$, is set to be $\sigma_l\simeq0.0064$ in order to evade the CMB $\mu$-distortion constraint\;\cite{Sasaki:2018dmp,Carr:2020gox}.

From the linear parametrization adopted in Eq.\,\ref{Eq:long_short_coupling} for $\eta<0$, the upper bound on \(\Phi_l\) should be imposed by requiring \(\nu \geq 0\), which implies \(\Phi_l \leq 1/|\eta|=\Phi_l^{\rm upp}\); and a lower bound on \(\beta\) in Eq.\,\ref{Eq:lower_bound_beta} determines the lower bound on \(\Phi_l=\Phi_l^{\rm low}\). Thus \(\Phi_l\) is bounded from below and above such that the corresponding cluster mass \(M_{\rm cl}\) is restricted to a finite range, over which the normalized PDF is given as 
\begin{align}\label{Eq:Pnorm}
P^{\rm norm}(M_{\rm cl})
=\frac{P(M_{\rm cl})\,
\Theta(M_{\rm cl}^{\rm upp}-M_{\rm cl})\,
\Theta(M_{\rm cl}-M_{\rm cl}^{\rm low})}
{\displaystyle \int_{M_{\rm cl}^{\rm low}}^{M_{\rm cl}^{\rm upp}} 
{\rm d}M_{\rm cl}\, P(M_{\rm cl}) }\,,
\end{align}
where $\Theta$ is the Heaviside step function, and \(M_{\rm cl}=M_{\rm cl}(\nu(\Phi_l))\) is given by Eq.\,\ref{Eq:MCl(Phi_l)} with \(M_{\rm cl}^{\rm low}=M_{\rm cl}\left(\nu(\Phi_l^{\rm low})\right)\) and \( M_{\rm cl}^{\rm upp}=M_{\rm cl}\left(\nu(\Phi_l^{\rm upp})=0\right)\) denoting the lower and upper bounds of the cluster mass, respectively. The SMBH mass function can then be expressed in terms of the probability function as 
\begin{align}\label{Eq:mass_function}
\Psi=n_{\rm tot}\,P^{\rm norm}(M_\bullet)\,M_\bullet\,\ln10\,,
\end{align}
where $n_{\rm tot}$ is the observed number density of SMBHs in a particular redshift range, and $M_\bullet\lesssim M_{\rm cl}$ is the SMBH mass resulting from the PBH cluster of mass $M_{\rm cl}$.

\subsection{The effect of accretion on the mass function} 
Accretion is crucial in modifying the initial mass function. Here we adopt the accretion model in Ref.\,\cite{Shapiro:2004ud}, assuming that accretion is dominated by baryonic matter. The mass accretion rate is given as \({\rm d}M/{\rm d}t=[\epsilon_L(1-\epsilon_M)/\epsilon_M]M/\tau\), where \(\tau\equiv Mc^2/L_E\simeq 0.45\,\mu_e^{-1}\,{\rm Gyr}\) is the characteristic accretion timescale with \(L_E\simeq 1.3\times 10^{46}\mu_e (M/10^8{\rm\,M}_\odot)\,{\rm erg}\,{\rm s}^{-1}\) the Eddington luminosity and \(\mu_e\) the mean molecular weight per electron. 
Here \(\epsilon_L\) denotes the Eddington luminosity ratio, and \(\epsilon_M\) is the radiative efficiency depending on the black-hole spin~\cite{Shapiro:2004ud}.

After integration, the enhancement of the black-hole mass is given by 
\begin{align}\label{Eq: accretion_disk_final_mass}
\frac{M_{f}(z)}{M_i}=\mathcal{F}_{\rm merg}\,\exp\left[\gamma_0\left(t_f(z)-t_i\right)\right],\quad t_f\leq t_{\rm max}
\end{align}
where \(\gamma_0\equiv \epsilon_L(1-\epsilon_M)/(\epsilon_M\tau)\), $\mathcal{F}_{\rm merg}$ is the enhancement factor due to mergers, $t_{i(f)}$ is the time at which the black hole has a mass $M_{i(f)}$, and $t_{\rm max}$ is the earliest time at which accretion can occur. The final time is given $t_f=t(z)$ as a function of redshift, which has an explicit form \(t(z)={2}/({3 H_0 \sqrt{\Omega_\Lambda}}) \ln\left[\left( {\Omega_\Lambda}/{\Omega_{\rm M}} \right)^{1/2}{(1 + z)^{-3/2}}+ \notag \right.
\\
\left.\sqrt{1 + \left(\Omega_\Lambda/\Omega_{\rm M}\right)(1 + z)^{-3}} \right]\), where the energy density of radiation is neglected since we focus on the late-time Universe when accretion is relevant.

We compute the mass function including the effects of accretion. The normalized mass function is defined as
\begin{align}
\Psi(M)\equiv M\frac{{\rm d}n}{{\rm d}M}\,,\quad 1=\int_{M_{\rm min}}^{M_{\rm max}}{\rm d}\ln M\, \Psi(M)\;.
\end{align}
Since accretion modifies the mass distribution to be a function of redshift, we define the redshift-dependent mass function as the fraction of PBHs in the mass interval \((M, M+{\rm d}M)\) at redshift \(z\),
\begin{align}
&\Psi(M_f(M,z),z)\,{\rm d}\ln M_f=\Psi(M,z_i)\,{\rm d}\ln M
\end{align}
where $M_f(M,z)$ is the final mass of a PBH with initial mass $M$ at redshift $z$ and $\Psi(M,z_i)$ is the initial mass function. Using the relation Eq.\,\ref{Eq: accretion_disk_final_mass}, the relation between the final mass after accretion and the initial mass is given by \({\rm d}\ln M/{\rm d}\ln M_f=1\). We then obtain the accretion-modified mass function 
\begin{align}
\Psi(M_f(M,z),z)&=\Psi(M(M_f,z),z_i)\;.
\end{align}

The accretion rates of SMBHs in the LRD phase are considered to be strongly accreting~\cite{Pacucci:2024tws,Inayoshi:2025hdr}, thus we set the accretion efficiency $\epsilon_L\sim \mathcal{O}(1)$. Since coherent accretion cannot last too long, we set the duty cycle of accretion as $f_{\rm duty}=0.01\textup{--}0.10$ and consider that accretion starts around $z\sim 20$.
 
\begin{figure}[H]
    \centering
    \includegraphics[width=0.49\textwidth]{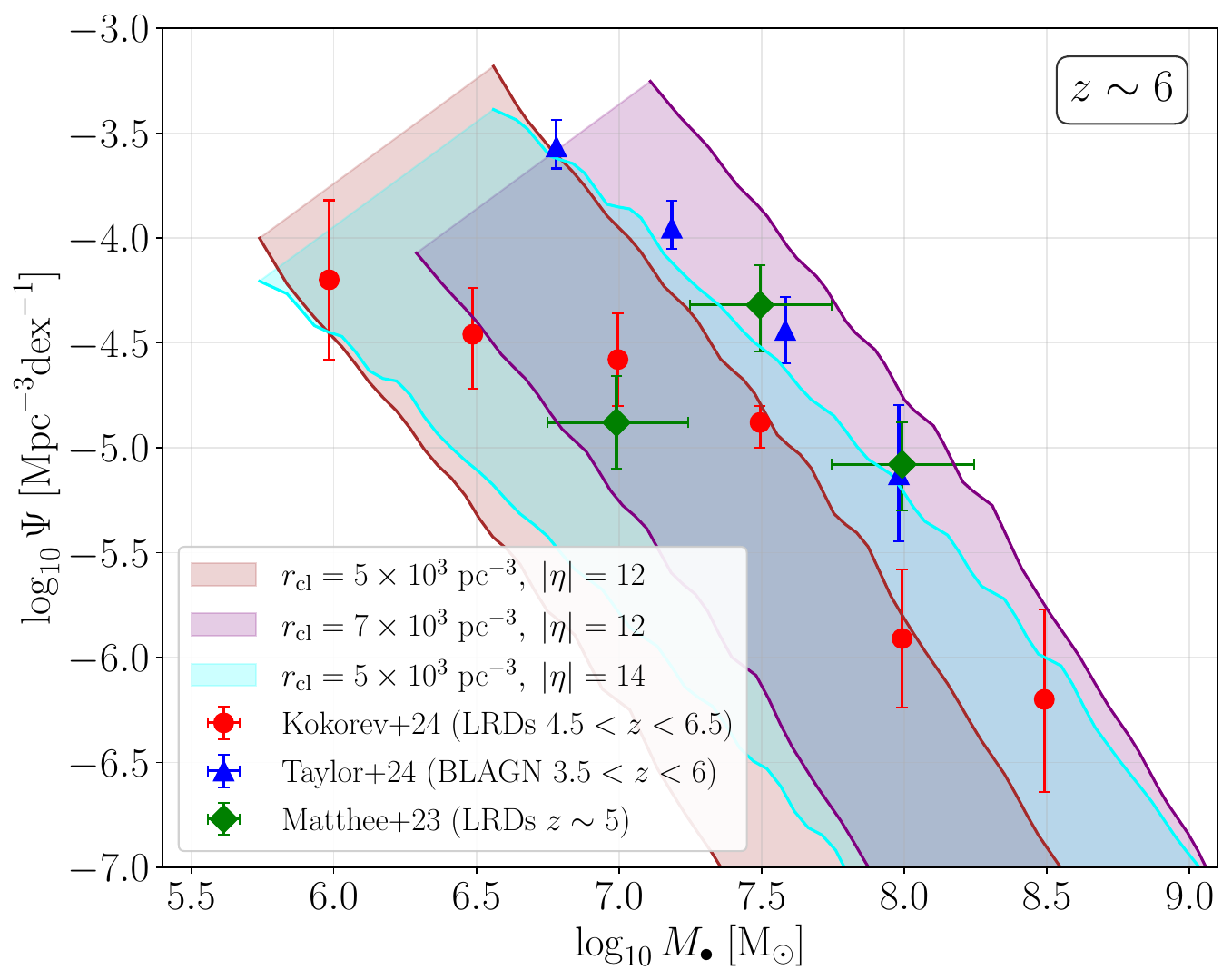}
    \caption{The mass function of SMBHs predicted by the PBH clusters through long--short mode coupling, Eq.\,\ref{Eq:long_short_coupling}.}
    \label{Fig:cluster_mass_function_accrete}
 \end{figure}

Fig.\,\ref{Fig:cluster_mass_function_accrete} shows the SMBH mass function predicted by the PBH-cluster scenario. Red points denote the LRDs identified in Ref.\,\cite{kokorev2024census} within $4.5<z<6.5$, blue triangles correspond to the broad-line AGN (BLAGN) samples from Ref.\,\cite{taylor2025broad} with an abundance roughly five times that of LRDs in $3.5<z<6$, and green diamonds mark the LRDs reported by Ref.\,\cite{matthee2024little} at $z\sim5$. We present results for clusters with different long–short mode couplings and long-mode length scales. The cyan band corresponds to $r_{\rm cl}=5\times 10^{3}\,\mathrm{pc}$ and $|\eta|=14$, the brown band to $r_{\rm cl}=5\times10^{3}\,\mathrm{pc}$ and $|\eta|=12$, and the purple band to $r_{\rm cl}=7\times10^{3}\,\mathrm{pc}$ and $|\eta|=12$. For all three models we impose a cutoff redshift $z_{\rm obs}\sim4$, since most observed LRDs lie above this redshift, where we adopt a total comoving number density of LRDs, $n_{\rm tot}=n_{\rm LRD}\simeq2\times10^{-5}\,\mathrm{cMpc}^{-3}$ from the observations~\cite{zhuang2025nexus}. In fact, our model predicts a much larger comoving number density of PBH clusters, \(n_\bullet\sim10^{-2}\textup{--}10^{-1}{\rm\,cMpc}^{-3}\), that can result in SMBHs~\cite{Zhang:2025tgm}. However, in Sec.\,\ref{Sec:discussion}, we estimate that due to the host halo properties and the possible major mergers, the observed PBH clusters in the form of LRDs are much reduced and consistent with observations.

For a given observed redshift, a larger long-mode scale $r_{\rm cl}$ raises the cutoff mass and enhances the abundance on the high-mass side; while a stronger coupling strength $|\eta|$ produces a shallower slope in the predicted mass function. Within the displayed parameter space, the model reproduces the observed high-redshift SMBH mass functions well. In Fig.\,\ref{Fig:cluster_mass_function_accrete}, the brown band matches well to the LRD mass function; and the purple band aligns closely with the BLAGN measurements. Consequently, this framework provides a viable parameter space capable of simultaneously accounting for the observed SMBH mass functions in both LRDs and BLAGNs.

%%%
%%%

\section{Seed effect and LRD properties}
\label{Sec:seed_effect}
From the previous section, for a suitable range of model parameters and a relatively low accretion duty cycle, SMBHs formed by runaway mergers of PBHs in clusters can reproduce the mass function inferred from JWST. Because our PBH clusters form on very small scales and persist as dense bound objects, it is tempting to explain some other properties of LRDs using the ``seed effect''~\cite{Carr:2018rid}.

The seed effect can naturally account for the overmassive feature of LRDs. It describes the mass that becomes bound to a heavy central seed (here, a dense PBH cluster) through gravity on small scales. Concretely, the halo bound to the cluster at redshift $z$ can be obtained as follows. First, the initial matter density contrast can be viewed as a linear perturbation, written as \( \delta_i(a_{\rm eq})={M_{\rm cl}}/{M_h}\), where \(M_h\) is the halo mass bound to the seed; and the density contrast grows \(\delta(z)=\delta_i({1+z_{\rm eq}})/({1+z})\propto (1+z)^{-1}\) during the matter-dominated era. As halos form when the density contrast exceeds the threshold \(\delta_c\simeq 1\), the halo mass bound to the cluster at redshift \(z\) is
\begin{align}\label{Eq:seed_halo}
M_h=M_{\rm cl}\frac{1+z_{\rm eq}}{1+z}\frac{1}{\delta_c}\simeq M_{\rm cl}\frac{1+z_{\rm eq}}{1+z}\;.
\end{align}
At around \(z\simeq 10\), a redshift probed by JWST, the bound halo mass is \(\mathcal{O}(10^2)\) times larger than the central cluster. 
\subsection{The stellar-to-BH ratio}
The associated stellar mass in the bound halo can be estimated as \(M_\star=\epsilon_\star f_b M_h\), where \(f_b\) is the fraction of baryons and \(\epsilon_\star\) is the star formation efficiency.
Hence, neglecting SMBH growth by accretion and approximating the SMBH mass by the original cluster mass, the stellar-to-BH mass ratio becomes
\begin{align}
\frac{M_\star}{M_\bullet}\simeq \epsilon_\star f_b\frac{1+z_{\rm eq}}{1+z}.
\end{align}
For reference values of the fraction of baryons $f_b\simeq 0.15$ and star formation efficiency $\epsilon_\star=0.1$ at $z\simeq10$, the stellar mass is $\sim\mathcal{O}(10^0\textup{--}10^1)$ times that of SMBHs, which explains the high stellar-to-BH mass ratio in LRDs. 

\begin{figure}[t]
    \centering
    \includegraphics[width=0.49\textwidth]{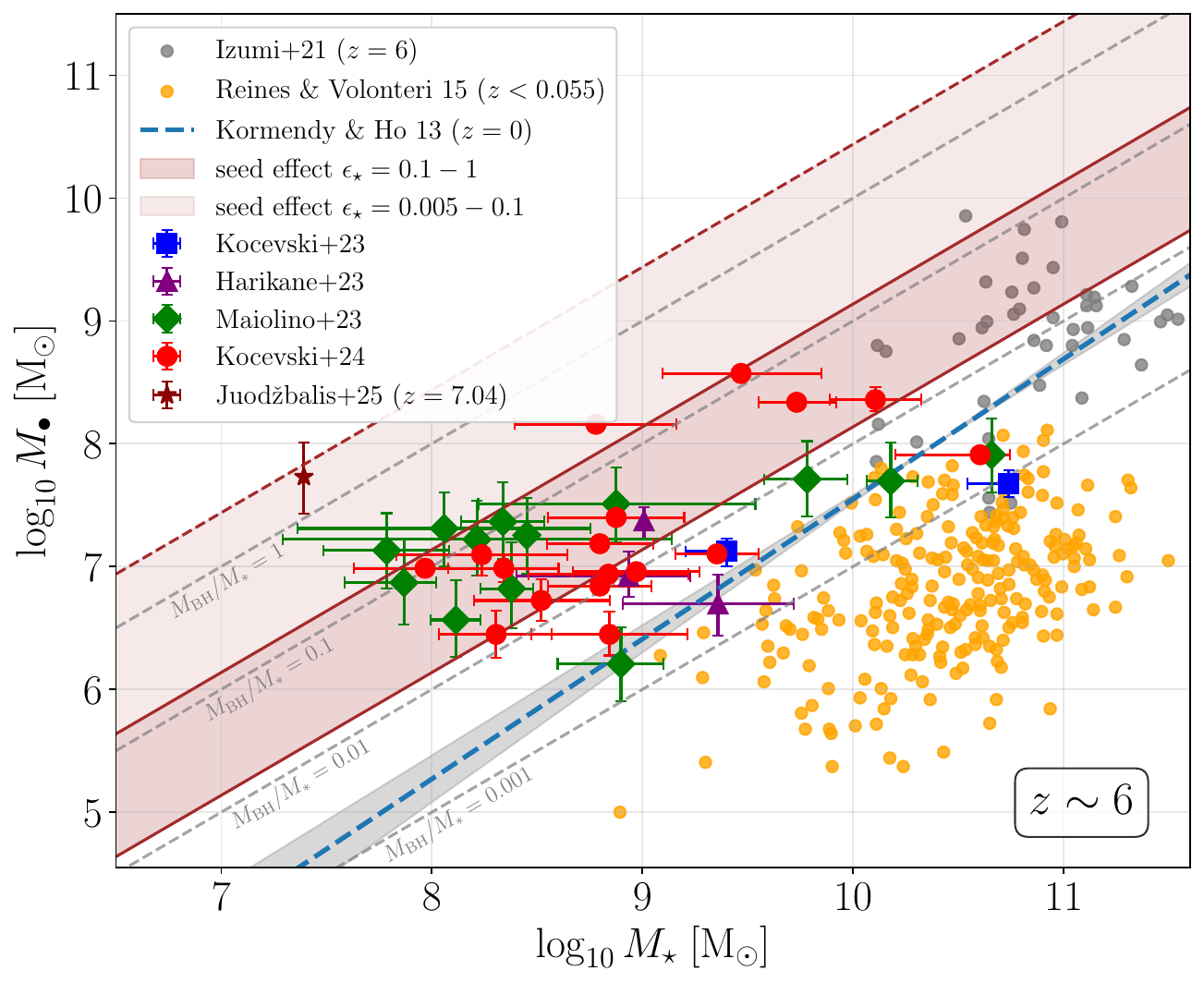}
    \caption{The stellar-to-BH ratio of halos induced by PBH clusters with different star formation rates.}
    \label{Fig:stellar-to-BH ratio}
 \end{figure}

Fig.\,\ref{Fig:stellar-to-BH ratio} shows the stellar-to-BH mass ratio in PBH cluster-induced halos. LRD observations from JWST at \(z>4\) are color-coded by their original sources~\cite{kocevski2023rise,kocevski2024rise,harikane2023jwst,maiolino2024jades,Maiolino:2025tih}. In particular, the brown star marks the recent detection of an almost naked SMBH reported in Ref.\,\cite{Maiolino:2025tih}. Dark-grey circles correspond to the \(z=6\) quasar samples~\cite{izumi2021subaru}; while orange points show the nearby BLAGN samples~\cite{reines2015relations}; and the blue dashed line is the best-fit local $M_\bullet\textup{--}M_\star$ relationship in massive
galaxies at $z=0$~\cite{kormendy2013coevolution}. 

The stellar-to-BH ratios in most LRDs can be naturally explained by the seed effect within the range of our theoretical predictions at \(z\sim6\) for different star-formation efficiencies shown in the dark-brown band for \(\epsilon_\star=0.1\textup{--}1\); and the light-brown band for \(\epsilon_\star=0.005\textup{--}0.1\). 

\subsection{The host halo spin and compactness of LRDs}
LRDs are observed as extremely compact objects, with galactic-disk effective radii of \(\sim100\,{\rm pc}\), much smaller than those of normal galaxies. The compactness of LRDs may result from low-spin haloes induced by PBH clusters through the seed effect. Moreover, our model can relate the central SMBHs to their host-halo properties, including the effective radius of the galactic disk and its redshift evolution.

\subsubsection{Halo spin distribution and the LRD effective radius} In our scenario, the compactness of LRD galaxies can be explained if they host PBH clusters embedded in low-spin haloes~\cite{mo1998formation,Pacucci:2025tyl}. The structural properties of disk galaxies are directly linked to the dimensionless spin parameter \(\lambda={J_h|E|^{{1}/{2}}}/({GM_h^{{5}/{2}}})\) of their host haloes, where $J_h$ is the the total angular momentum of the halo with mass $M_h$, and $E$ is its total energy~\cite{Peebles:1969jm}.

The distributions of $\lambda$ found in various cosmological $N$-body simulations~\cite{barnes1987angular,warren1992dark,Cole:1995ep,lemson1999environmental} depend only very weakly on halo mass and the form of the initial spectrum of density fluctuations. From the simulations, $\lambda$ follows a lognormal distribution,
\begin{align}\label{Eq:P(lambda)}
P(\lambda)=\frac{1}{\lambda\sqrt{2\pi}\,\sigma_{\ln\lambda}}\exp\left[-\frac{\left(\ln\lambda/\bar{\lambda}\right)^2}{2\sigma_{\ln\lambda}^2}\right]
\end{align}
which is nearly independent of halo mass and redshift, where the value of the median spin and variance is given by $\bar{\lambda}=0.05$ and $\sigma_{\ln\lambda}=0.5$~\cite{mo1998formation}. Since \(\lambda\) is insensitive to the initial power spectrum of perturbations, and PBH clusters enhance the matter power spectrum only on small scales, the spin distribution Eq.\,\ref{Eq:P(lambda)} should remain valid in our PBH clustering model. 

Infalling baryons in halos will settle into a rotationally supported disk at the halo center. The scale length $r_d$ of the disk is associated with the size and the spin of the host halo, given by \(r_d=(1/\sqrt{2})\left({j_d}/{m_d}\right)\lambda\,r_{200}\), where $j_d=J_d/J_h$ is the fraction of the halo angular momentum acquired by the disk, $m_d=M_d/M_h$ is the baryon fraction that settles into the disk, and $r_{200}(M_h,z)=\left({GM_{h}}/{100H^2(z)}\right)^{{1}/{3}}$ is the virial radius of the halo within which the mean mass density is $200$ times the critical density of the background~\cite{Gunn:1972sv} with $H(z)\simeq H_0\sqrt{\Omega_{\rm M}\left(1+z\right)^3+\Omega_\Lambda}$ the Hubble parameter. At high redshifts, most of the available baryons are likely to form galaxies, so the spin distribution of dark-matter halos can be directly related to that of galaxies. We assume that baryons retain their specific angular momentum during collapse, so the gas disk will form without significant loss of angular momentum, 
i.e., $j_d/m_d=1$. Together with identifying $r_{\rm eff}=r_d$, we can determine the effective radius \(r_{\rm eff}=(1/\sqrt{2})\,\lambda\,r_{200}(M_h,z)\). Thus, the critical spin parameter of LRDs required to produce a galaxy with compactness $r_{\rm eff}$ is \(\lambda_{\rm LRD}(z)={\sqrt{2}\,r_{\rm eff}}/{r_{200}(M_h,z)}\). Moreover, with Eq.\,\ref{Eq:P(lambda)} the fraction of haloes that can produce the galaxies more compact than the LRD effective size can be computed by 
\begin{align}\label{Eq:F(lambda)}
F(\lambda<\lambda_{\rm LRD})=\int^{\lambda_{\rm LRD}(z)}_0P(\lambda){\rm\,d}\lambda\;,
\end{align}
which is crucial in estimating the LRD abundance.

For the seed-induced halo mass, Eq.\,\ref{Eq:seed_halo}, bounded by a PBH cluster in our modeling, the virial radius is then
\begin{align}\label{Eq:r200_seed_effect}
r_{200}^{\rm seed}=\left[\frac{G M_{\rm cl}(1+z_{\rm eq})}{100(1+z)H^2(z)}\right]^{1/3}\;.
\end{align}
The seed effect relates the cluster mass to the halo radius and thus the effective radius observed in LRDs given a $\lambda<\lambda_{\rm LRD}$. We note that for a fixed cluster mass in the matter era, the virial radius $r_{200}^{\rm seed}\propto(1+z)^{-4/3}$.

\subsubsection{Compactness of LRDs} 
\begin{figure}[htbp]
    \centering
    \includegraphics[width=0.49\textwidth]{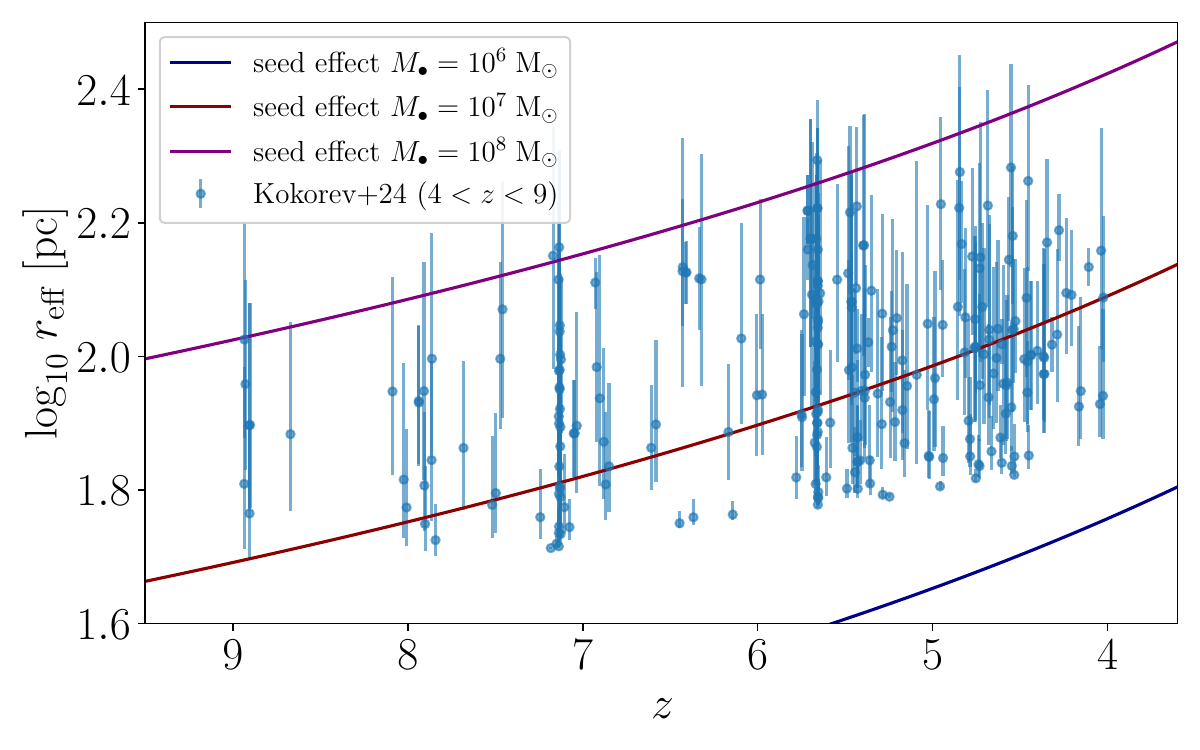}
    \caption{The effective radius of galaxies in haloes induced by PBH clusters for various cluster masses $\sim10^6\textup{--}10^8\,\mathrm{M}_\odot$ over a range of redshifts, $4<z<9$. 
}\label{Fig:Reffseed_z}
 \end{figure}

If LRDs are associated with halos of low angular momentum $\lambda_{\rm LRD}\lesssim \mathcal{O}(10^{-2})$, we can obtain the effective radius \(r_{\rm eff}=(1/\sqrt{2})\lambda_{\rm LRD}\, r_{200}^{\rm seed}\) of the galaxies in halos induced by the heavy seed. At redshift $z=5$, $r_{\rm eff}\simeq46\textup{--}215\, \text{pc}$ corresponding to cluster masses (subsequent SMBH masses) $M_\bullet\approx M_{\rm cl}=10^6\textup{--}10^8{\rm\,M}_{\odot}$ with \(\lambda\sim 0.015\) (see Fig.\,\ref{Fig:Reffseed_z}). In Sec.\,\ref{Sec:discussion}, we show that this low-spin value is also consistent with the observed abundance of the LRDs.

Fig.\,\ref{Fig:Reffseed_z} shows the effective radius of galaxies in haloes induced by PBH clusters for various cluster masses over a range of redshifts. Blue points are taken from Ref.\,\cite{kokorev_census_2024}; each marker denotes the median (50th percentile) physical effective radius, and the error bars indicate the 16th and 84th percentiles. The dark-blue, dark-red, and purple solid lines correspond to the effective radii of galactic disks in seed-induced haloes with SMBH masses $10^6$, $10^7$, and $10^8{\rm\,M}_\odot$, respectively, resulting from the initial PBH clusters. With \(\lambda\sim 0.015\), the mass range $10^6\textup{--}10^8\,\mathrm{M}_\odot$ reproduces the data well, consistent with the observed masses of LRDs. Moreover, the model successfully captures the observed trend of increasing effective radius with decreasing redshift.

To translate SMBHs in our model into observations, we assign a bright rest-frame ultraviolet (UV) luminosity $M_{\rm UV}$ to each black hole by assuming a population-averaged Eddington ratio, \(L_{\rm bol}=f_{\rm Edd}L_{\rm Edd}\), with the standard electron-scattering Eddington luminosity \(L_{\rm Edd}=1.26\times10^{38}(M_\bullet/{\rm M}_\odot)\,{\rm erg\,s^{-1}}\)~\cite{RybickiLightman1979}. We then convert \(L_{\rm bol}\) to \(L_\nu\) at 1450\,\(\text{\AA}\) using \({\rm BC}_{1450}\equiv L_{\rm bol}/(\nu L_\nu)=4.0\), consistent with empirical quasar bolometric corrections \citep{Runnoe2012}, and express \(L_\nu\) as an absolute AB magnitude using the standard AB zero point \citep{OkeGunn1983,Hogg1999}.  This conversion assumes isotropic emission and fixed population-averaged \(f_{\rm Edd}\) and \({\rm BC}_{1450}\).

\begin{figure}[htbp]
    \centering
    \includegraphics[width=0.49\textwidth]{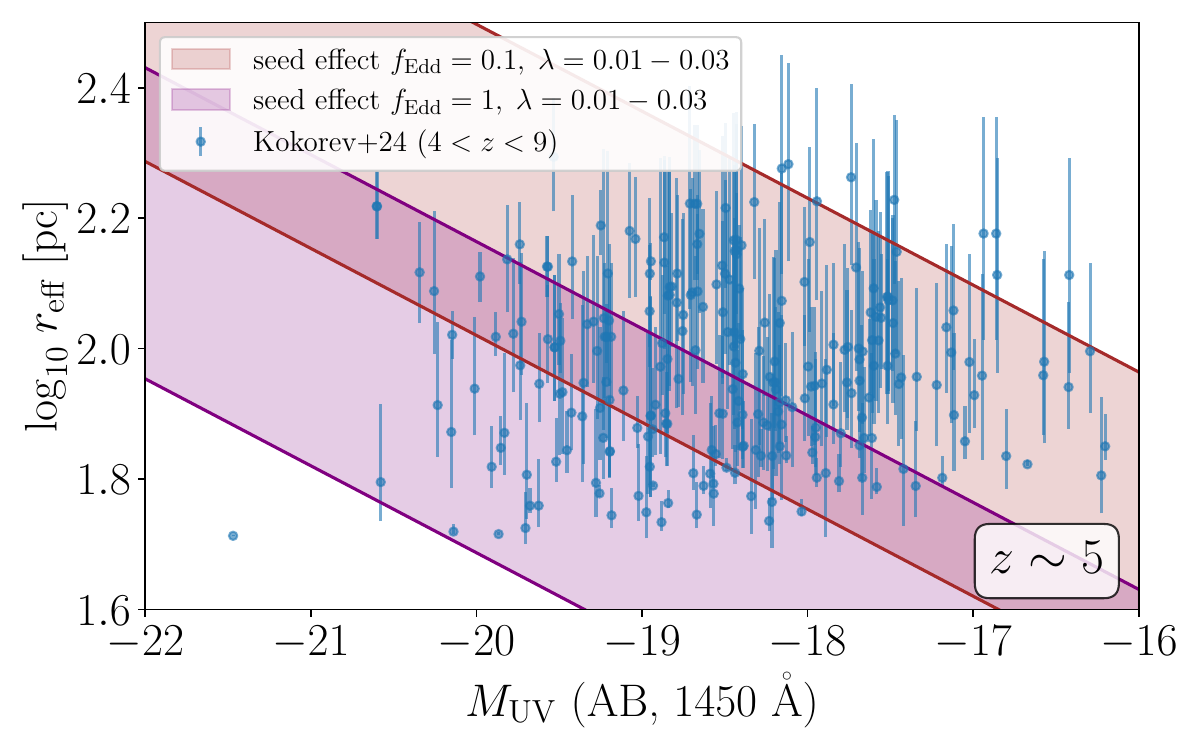}
    \caption{The effective radius $r_{\rm eff}$ of galaxies in haloes induced by the seed effect of SMBHs formed in PBH clusters as a function of the UV luminosity $M_{\rm UV}$. }\label{Fig:Reffseed_Muv}
 \end{figure}

Fig.\,\ref{Fig:Reffseed_Muv} presents the relation between the effective radius $r_{\rm eff}$ and the UV luminosity magnitude $M_{\rm UV}$. Blue points are taken from Ref.\,\cite{kokorev_census_2024} similarly as in Fig.\,\ref{Fig:Reffseed_z}. The brown band corresponds to SMBHs accreting at $f_{\rm Edd}=0.1$, and the purple band to $f_{\rm Edd}=1$. Both bands assume low-angular-momentum haloes with spin parameter $\lambda=0.01$--$0.03$ at redshift $z=5$. The two bands, together with the region between them, encompass the majority of the observed effective radii.

 
\section{The dense gas in LRDs} 
\label{Sec:LRD_dense_gas}
Dense gas is a key feature of LRDs and can account for several of their observed properties~\cite{inayoshi2025extremely,de2025remarkable,Kido:2025vhx,rusakov2026little,kokorev2026deepest}. LRD spectra often exhibit strong absorption features superposed on broad Balmer emission lines, together with a spectral break near the Balmer limit. These characteristics require a high covering fraction of dense gas around the SMBH; and atomic hydrogen number densities of $\gtrsim10^{9}\text{--}10^{11}\,\mathrm{cm}^{-3}$ can reproduce them~\cite{inayoshi2025extremely}. Moreover, if the accreting BH is embedded in a dense, quasi-spherical envelope, the weakness of X-ray emission and of radio jets can be naturally explained by strong absorption and jet confinement in such an environment~\cite{Kido:2025vhx}. 

While the detailed gas and accretion dynamics and the resulting density profile in PBH clusters are beyond the current scope, here we demonstrate that the observed dense gas is consistent with our modeling given the accretion disk model in the existing literature.

In our scenario, the host haloes of LRDs having low angular momentum tend to develop a compact baryonic concentration. As halos possess an angular momentum~\cite{Jiang:2018ioo}, infalling baryons are likely to conserve their specific angular momentum and settle into a rotationally supported disk at the halo center, producing the required high central gas density. The central number density of gas in the fat, cold disks of high-redshift haloes may be written as~\cite{mo1998formation,oh2002second,Volonteri:2005fj}:
\begin{align}\label{Eq:ngas}
n_0 &\simeq 6\times 10^4\left(\frac{f_d}{0.5}\right)^2\left(\frac{\lambda}{0.05}\right)^{-4}\left(\frac{T_{\rm gas}}{8000\ \mathrm{K}}\right)^{-1} \notag\\
&\qquad\times\left(\frac{M_h}{10^9\,\mathrm{M}_\odot}\right)^2\left(\frac{r_{200}^{\rm seed}}{6{\rm \,kpc}}\right)^{-4}\,\mathrm{cm}^{-3}\,,
\end{align}
where \(f_d\) is the fraction of gas that settles in the disk, \(\lambda\) is the spin parameter, \(T_{\rm gas}\) is the gas temperature, and \(M_h\)  and \(r_{200}^{\rm seed}\) are given by Eqs.\,\ref{Eq:seed_halo} and \ref{Eq:r200_seed_effect}, respectively. The gas number density at the scale of the galactic disk scales inversely with the fourth power of the spin parameter \(\lambda\). For a low-spin halo of \(\lambda\lesssim 0.01\), the central density is enhanced by \(\gtrsim\mathcal{O}(10^3)\) compared to that of $\lambda=0.05$.

Owing to the seed effect, dense PBH clusters can induce compact dark-matter plus gas overdensities when decoupling from the Hubble flow before matter–radiation equality. They can seed deep, centrally concentrated potentials earlier than in the standard picture, where baryons fall in only after a dark-matter halo has formed. In the matter-dominated era, these seed potentials accelerate the infall of baryons and collisionless matter at very high redshifts; therefore, seed-induced haloes are expected to retain a larger baryon abundance than ordinary haloes. Specifically, a cluster's gravity draws in collisionless matter and baryons, and under secondary-infall/adiabatic-growth conditions~\cite{Carr:1984ikr,Bertschinger:1985pd,Carr:2018rid} the inner density profile of a seed-induced halo can be substantially steeper than a standard Navarro–Frenk–White cusp~\cite{Navarro:1996gj}.

We do not expect such a steep power-law profile to persist indefinitely, because complex baryonic processes near the black hole and subsequent merger activity can modify the halo structure and erase memory of the initial conditions. Nevertheless, it is reasonable to assume that the gas density in the inner region of a seed-induced halo, which is close to the central black hole, may be enhanced relative to a typical halo. This enhancement provides a plausible explanation for the dense gas observed in LRDs.

In our model, prior to merging into a single object, a PBH cluster behaves as a collection of compact objects rather than as a single massive accretor. Assume Bondi–Hoyle–Lyttleton gas accretion at an early stage. Two physical effects render coherent, cluster-scale accretion inefficient: (i) the Bondi capture radius of individual PBHs is typically much smaller than the mean PBH separation, and (ii) PBHs acquire large relative velocities after virialization, which strongly suppress accretion. The Bondi accretion rate for a single PBH in a cluster is roughly $\propto(Gm_{\rm pbh})^2\rho_{\rm gas}^{\rm cl}/(c_s^2+v^2)^{3/2}$, where \(\rho_{\rm gas}^{\rm cl}\) is the gas mass density within the cluster, and the Bondi radius is estimated to be $\sim Gm_{\rm pbh}/(c_s^2+v^2)$ with $c_s$ the sound speed of the gas and $v$ the velocity dispersion of PBHs. 

Typically, for single PBHs with mass $m_{\rm pbh}=30{\rm\,M}_\odot$ in a $N_{\rm cl}=10^5$ cluster of radius $\simeq0.05{\rm\,pc}$~\cite{Zhang:2025tgm}, the averaged separation of PBHs $n_{\rm cl}^{-1/3}\sim 10^{-3}{\rm\,pc}$ is much larger than the individual Bondi radii $\sim 10^{-6}{\rm\,pc}$ for $c_s\lesssim v\simeq500{\rm\,km/s}$. Therefore, the Bondi radii of single PBHs do not overlap and they accrete independently from smaller regions. The total accretion rate $\sim N_{\rm cl}(Gm_{\rm pbh})^2\rho_{\rm gas}/(c_s^2+v^2)^{3/2}$~\cite{Kaaz:2019wdi} is much smaller than the single-SMBH accretion rate $\sim N_{\rm cl}^2(Gm_{\rm pbh})^2\rho_{\rm gas}/(c_s^2+v_\bullet^2)^{3/2}$ given the same total mass $N_{\rm cl}\,m_{\rm pbh}$, where $v_\bullet$ is the peculiar velocity of a single SMBH. The latter is also the rate of gas inflow by the accretion of the whole cluster as a single object given \(\rho_{\rm gas}=\rho_{\rm gas}^{\rm disk}\simeq n_0 m_{\rm H}\) the hydrogen gas mass density in the galatic disk outside the cluster, with $n_0$ estimated by Eq.\,\ref{Eq:ngas} and \(m_{\rm H}\) the hydrogen mass. The ratio of these two accretion rates $1/N_{\rm cl}\times [(c_s^2+v_\bullet^2)/(c_s^2+v^2)]$ is suppressed by the total number $N_{\rm cl}$ if $v_\bullet\ll v\sim c_s$. 

Therefore, the gas density within the cluster region must be $\rho_{\rm gas}^{\rm cl}\simeq N_{\rm cl}\,\rho_{\rm gas}^{\rm disk}$ when the gas consumption of the small PBHs is balanced by the gas inflow outside the cluster. Alternatively, the gas number density can be estimated as $N_{\rm cl}\,n_0 \simeq \mathcal{O}(10^{11}){\rm\,cm}^{-3}$ taking $N_{\rm cl}=10^5,\;\lambda\simeq \mathcal{O}(10^{-2})$ and other parameters as fiducial, which is consistent with the dense gas environment around SMBHs in LRDs. As a result, a large amount of gas can accumulate in the vicinity of clusters;
%this may naturally explain the dense gaseous envelopes observed in LRDs. 
after single SMBHs emerge, the rate of consuming gas increases dramatically, allowing luminous accretion to proceed on the existing gas reservoir, thereby accounting for the strong accretion phase of LRD SMBHs.

\section{Accelerating early massive galaxy formation with SMBHs from PBH clusters}
\label{Sec:high-z_galaxy}
JWST has revealed a population of unexpectedly massive and luminous galaxies at very high redshift, which may be in tension with early structure formation in the standard $\Lambda$CDM model. In particular, the red massive galaxy candidates reported at $z\gtrsim 7$ imply stellar masses and number densities that are difficult to accommodate with the abundance of massive dark-matter halos available at such early times~\cite{Labbe:2022ahb}. This tension is related to the required baryon-to-star conversion efficiency. If the inferred stellar masses are correct, some systems appear to require extremely rapid star formation, with efficiencies approaching or even exceeding the available baryonic budget in host halos~\cite{Boylan-Kolchin:2022kae}. For example, the cumulative stellar-mass density at $z\simeq 10$, $\rho_\star(>10^{10}{\rm\,M}_\odot)\simeq 1.3^{+1.1}_{-0.6}\times10^6{\rm\,M}_\odot\,{\rm Mpc}^{-3}$, already lies close to the maximal expectation from standard hierarchical assembly.

However, the interpretation of this high-redshift galaxy tension remains open~\cite{Parashari:2023cui}. The inferred stellar masses and abundances depend on stellar-population modeling, dust attenuation, AGN contamination, sample variance, and Eddington bias~\cite{Krishnan:2025dti}. Some of the apparent discrepancy may be reduced by accounting for these systematics, by allowing for burstier and more efficient early star formation, or by including moderate AGN contributions~\cite{Sarkar:2026hie}. A consensus is yet to be reached, but these studies strongly suggest that galaxy formation in the first few hundred million years was more rapid and efficient than predicted by many pre-JWST models.

In our model, we can also account for the appearance of massive, high-redshift galaxies. Besides the SMBHs that give rise to LRDs, the population contains many other, massive SMBHs that do not evolve into LRDs; these massive seeds can accelerate early galaxy formation via the seed effect. In particular, at very high redshifts (\(z\gtrsim 10\mbox{--}12\)) the high-mass end (\(M_\bullet\gtrsim 10^{8}\,\mathrm{M}_\odot\)) is especially important for driving rapid baryon infall and early stellar assembly.

Following Refs.\,\cite{Fakhouri:2010st,Hai-LongHuang:2024gtx,inayoshi2022lower}, we assume that linear growth of the density contrast ceases once perturbations enter the non-linear regime by \(z_{\rm cut}\). After this redshift (i.e. for \(z<z_{\rm cut}\)), halo growth proceeds through accretion and mergers and can be described by an analytic form calibrated to cosmological \(N\)-body simulations~\cite{Fakhouri:2010st,inayoshi2022lower}: 
\begin{align}\label{Eq:Mh_growth_nbody}
\langle\dot{M}_h&\rangle\simeq 46.1\mathrm{M}_\odot\text{yr}^{-1}\left(\frac{M_h}{10^{12}\mathrm{M}_\odot}\right)^{1.1}(1+1.11z)\notag\\
&\times\sqrt{\Omega_{\rm M}(1+z)^3+\Omega_\Lambda}\;.
\end{align}
Upon integration, we obtain
\begin{align}\label{Eq:mass_grow_halo}
& (M_h/\mathrm{M}_\odot)(z)\notag\\
&=  \left[(M_h(z_{\rm cut})/\mathrm{M}_\odot)^{-0.1}\right.\notag\\
&\left.+1.11\textbf{C}_1(z-z_{\rm cut})+0.11\textbf{C}_1\mathrm{log}\left(\frac{1+z_{\rm cut}}{1+z}\right)\right]^{-10}
\end{align}
where $\textbf{C}_1=0.004213$ and the halo mass at $z_{\rm cut}$ is given by the seed effect \( M_h(z_{\rm cut})=M_\bullet({1+z_{\rm eq}})/({1+z_{\rm cut}})\). The cumulative stellar mass density of haloes formed through the seed effect is computed using the mass function from the cluster model, Eq.\,\ref{Eq:mass_function},
\begin{align}
\rho_{\star}(>M_{\star})
&=\epsilon_{\star}f_b\int^{\infty}_{\ln\frac{M_{\star}}{\epsilon_\star f_b}}\frac{{\rm d}n}{{\rm d}\ln M_h}M_h{\rm\,d}\ln M_h\\
=\epsilon_{\star}f_b&\int^{\infty}_{\ln M_\bullet\left(\frac{M_{\star}}{\epsilon_\star f_b},z,z_\mathrm{cut}\right)}n_{\rm tot}\,P^{\rm norm}(M_\bullet)\notag\\
\times M_h&(M_\bullet,z,z_\mathrm{cut})\,M_\bullet{\rm\,d}\ln M_\bullet\;.
\end{align}

While in the \(\Lambda\)CDM model, the cumulative stellar mass density above a stellar mass threshold \(M_{\star}\) at redshift \(z\) is associated with the Press–Schechter halo mass function (with the linear matter power spectrum supplied by \textbf{CLASS}~\cite{Lesgourgues:2011re}), given by
\begin{align}
\rho_{\star}(>M_{\star})&=\epsilon_{\star}f_b\int^{\infty}_{\ln\frac{M_{\star}}{\epsilon_\star f_b}}\frac{{\rm d}n}{{\rm d}\ln M_h}\bigg|_{\rm PS}M_h{\rm\,d}\ln M_h
\end{align}
where the mass function reads \({{\rm d}n(M,z)}/{{\rm d}\ln M}\big|_{\rm PS}=(\bar{\rho}_{\rm M}/M)f(\sigma_M,z)\big|{{\rm d}\ln \sigma_M^{-1}(M)}/{{\rm d}\ln M}\big|\) with \(f(\sigma_M,z)=\left({2}/{\pi}\right)^{{1}/{2}}{\delta_c(z)}/{\sigma_M}\exp\left(-{\delta^2_c(z)}/{2\sigma_M^2}\right)\) the multiplicity function, where \(\bar{\rho}_{\rm M}\simeq3.95\times10^{10}{\rm\,M}_\odot{\rm\,Mpc}^{-3}\) is the mean matter density today, \(\delta_c(z)\simeq 1.686/D(z)\) is the linear collapse threshold at redshift \(z\) with \(D(z)\) the linear growth factor~\cite{Hamilton:2000tk}, and \(\sigma_M\) is the variance of the linear density field on the mass scale \(M\) at $z=0$. 

\begin{figure}[htbp]
    \centering
    \includegraphics[width=0.49\textwidth]{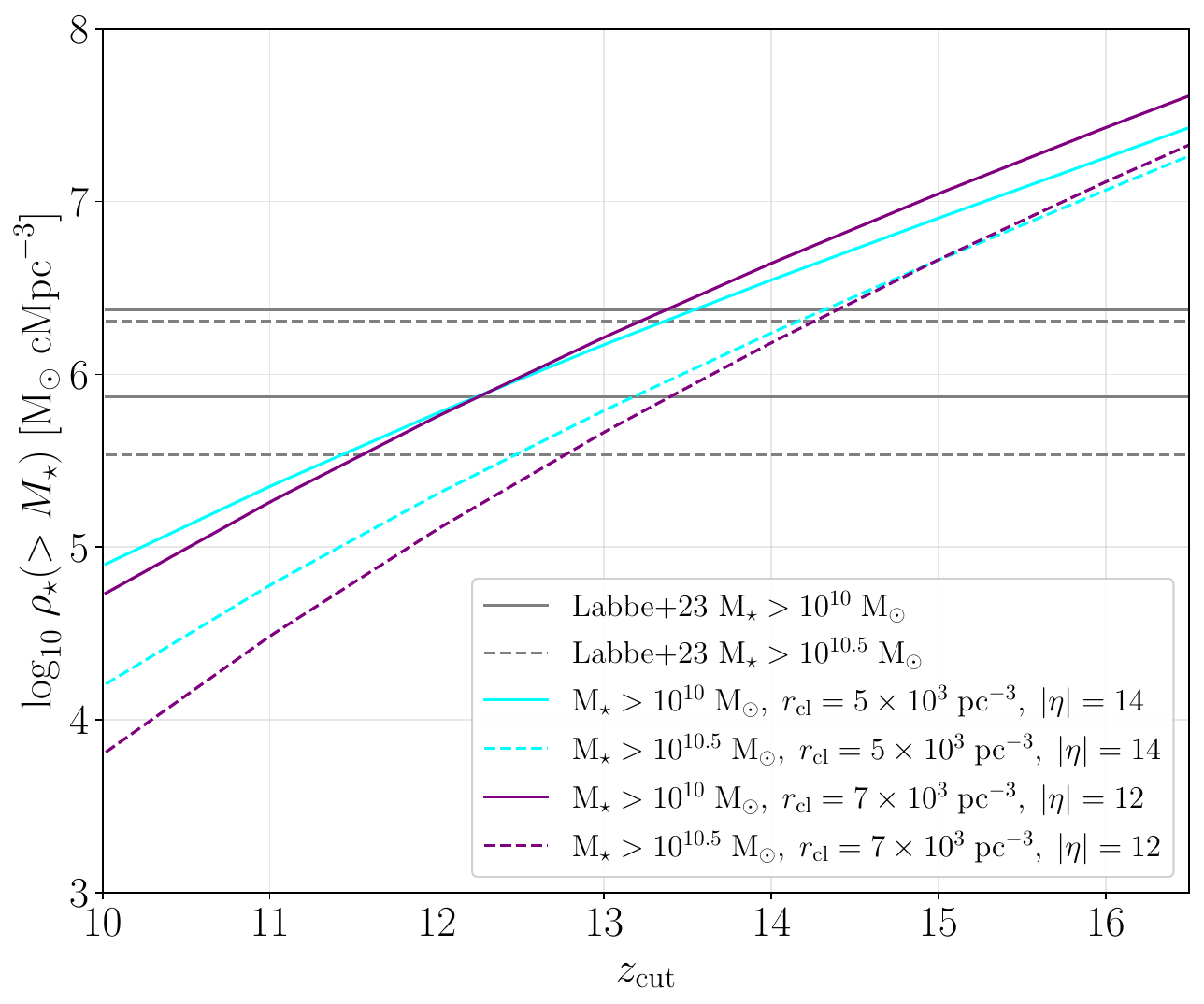}
    \caption{The cumulative stellar mass density of haloes formed through the seed effect at \(z\simeq10\) with different cutoff redshifts.
    }\label{Fig: The cumulative stellar mass density zcut}
 \end{figure}

Fig.\,\ref{Fig: The cumulative stellar mass density zcut} shows the cumulative stellar-mass density of haloes formed via the seed effect for various \(z_{\rm cut}\) in the validity range, where we set the star-formation efficiency \(\epsilon_\star=1\) and regard model curves that lie above the lower observational error bar as capable of explaining the measurement. In the plot, we adopt the fixed parameters \(\nu_g=8.5\) and \(\kappa=0.1313\), for which the PBH abundance is \(f_{\rm pbh}\sim 10^{-5}\). The total comoving number density of SMBHs is then estimated as \(n_{\bullet}\simeq f_{\rm pbh}\bar{\rho}_{\rm M}/M_\bullet\simeq 10^{-1}\,{\rm Mpc}^{-3}\) with \(M_\bullet=10^6{\rm\,M}_\odot\). The horizontal grey-solid line marks the JWST measurement~\cite{Labbe:2022ahb} at \(z\simeq 10\) for \(\rho_\star(\gtrsim 10^{10}{\rm\,M}_\odot)\simeq 1.3^{+1.1}_{-0.6}\times 10^{6}{\rm\,M}_\odot\,\mathrm{Mpc}^{-3}\), while the grey-dashed line indicates the measurement for \(\rho_\star(\gtrsim 10^{10.5}{\rm\,M}_\odot)\simeq 1.3^{+11}_{-6}\times 10^{5}{\rm\,M}_\odot{\rm\,Mpc}^{-3}\). The cyan and purple curves correspond to the parameter choices \((|\eta|, r_{\rm cl})=(14,5\times10^{3}{\rm\,pc})\) and \((|\eta|, r_{\rm cl})=(12,7\times10^{3}{\rm\,pc})\), respectively.

To explain the JWST observations requires a cutoff redshift of roughly \(z\gtrsim 13\). This cutoff redshift marks the epoch at which the halo-growth formula calibrated by \(N\)-body simulations takes over the growth driven by the seed effect, because the assumption that the induced halo remains isolated is no longer valid. We estimate this redshift as the time when the comoving number density of dark halos, computed from the Press--Schechter formula in the \(\Lambda\)CDM model for halo masses corresponding to the seed-effect-induced mass with seed masses of \(10^{6\textup{--}7}{\rm\,M}_\odot\), becomes comparable to the comoving number density of PBH clusters, \(n_{\rm tot}=n_{\bullet}\sim 0.1\,{\rm Mpc}^{-3}\). This criterion gives \(z\lesssim 16\). Therefore, the cutoff redshift should be in the range \(z\in[13,16]\). Here we adopt the fiducial cutoff redshift \(z_{\rm cut}=15\) as it is the highest redshift at which Eq.\,\ref{Eq:Mh_growth_nbody} is valid from  simulations~\cite{Fakhouri:2010st}.

\begin{figure}[htbp]
    \centering
    \includegraphics[width=0.49\textwidth]{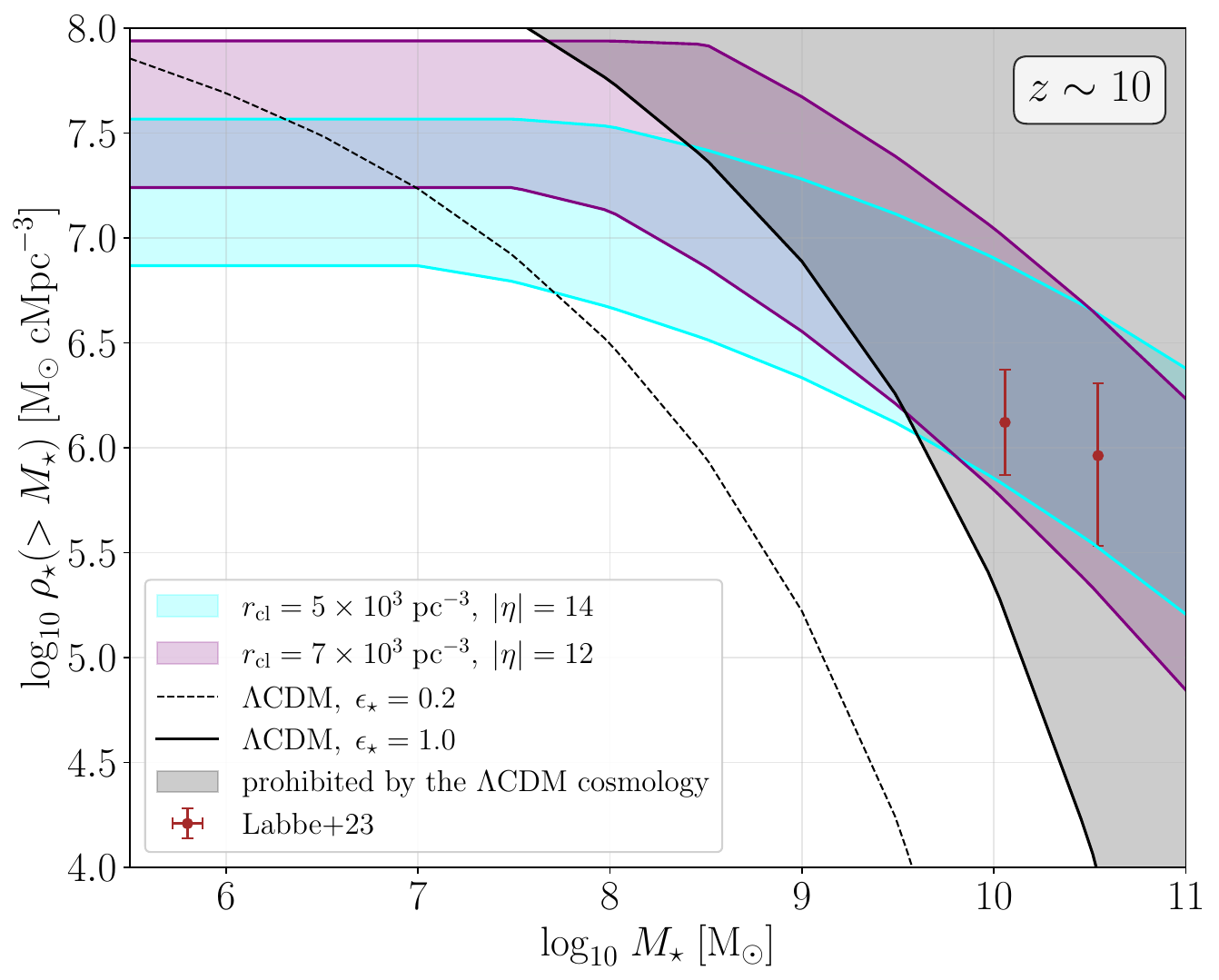}
    \caption{The cumulative stellar mass density with different $M_\star$ at redshift $z=10$. 
    }\label{Fig: The cumulative stellar mass density atz=10}
 \end{figure}

In Fig.\,\ref{Fig: The cumulative stellar mass density atz=10}, to illustrate the dependence on \(M_\star\), we show results for two parameter sets at redshifts \(z=10\)~\cite{Labbe:2022ahb}. The cyan band corresponds to \(|\eta|=14,\; r_{\rm cl}=5\times 10^{3}{\rm\,pc}\), and the purple band to \(|\eta|=12,\; r_{\rm cl}=7\times10^{3}{\rm\,pc}\), and the star formation efficiencies are both set as \(\epsilon_\star=0.2\textup{--}1.0\). Black curves show the \(\Lambda\)CDM prediction, with \(\epsilon_\star=0.2\) for the thin dashed line and \(\epsilon_\star=1.0\) for the thick solid line. The grey region above the black solid line is unphysical within \(\Lambda\)CDM because there the stellar mass would exceed the total baryon mass available in the Universe. Since some observed values lie within this excluded region, \(\Lambda\)CDM alone struggles to explain the measurements. By contrast, with appropriate parameter choices our model—via the early appearance of heavy SMBH seeds that accelerate galaxy formation—can reproduce the observed cumulative stellar-mass density and thus help alleviate the tension between the observations and the standard \(\Lambda\)CDM prediction.

\section{Discussion}
\label{Sec:discussion}
The abundance of SMBHs predicted in our PBH cluster model is broadly consistent with the observed LRDs. Furthermore, LRDs may gradually evolve into ordinary AGNs in the local Universe, thereby approaching the SMBH abundance predicted by our model.

\subsection{Estimation of LRDs abundance} In our model the PBH abundance is approximately \(f_{\rm pbh}\sim10^{-6}\textup{--}10^{-5}\) \cite{Zhang:2025tgm}, so the comoving number density of SMBHs generated via runaway mergers of PBH clusters can be estimated as \(n_\bullet\simeq f_{\rm pbh}\,\bar{\rho}_{\rm M}/M_\bullet\sim10^{-2}\textup{--}10^{-1}{\rm\,cMpc}^{-3}\), where we have adopted a characteristic pristine SMBH seed mass \(M_\bullet\sim10^{6}{\rm\,M}_\odot\) with $\bar{\rho}_{\rm M}$ the present-day mean matter density. We now show that, when compactness, major mergers, and duty cycle are taken into account, the SMBH abundance generated from PBH clusters can be consistent with LRD observations. 

The compactness of LRDs is related to the spin of the host halos. As discussed, LRDs are associated with low-spin haloes that satisfy \(\lambda\lesssim \mathcal{O}(10^{-2})\) to match their observed compactness. These haloes account for a fraction \(F(\lambda<\lambda_{\rm LRD})\sim 10^{-2}\) of the seed-induced halo population according to Eq.\,\ref{Eq:F(lambda)}. Thus, LRDs are embedded in roughly \(1\%\) of the non-merged seed-induced haloes and \(n_{\rm compact}=F(\lambda<\lambda_{\rm LRD})\,n_{\rm seed}\sim10^{-4}\textup{--}10^{-3}{\rm\,cMpc}^{-3}\), where \(n_{\rm seed}\) denotes the comoving number density of seed clusters (i.e. \(n_{\rm seed}\sim n_{\bullet}\)).
 
Apart from compactness, the SMBHs embedded in LRDs are overmassive. Major mergers between halos can substantially modify these features and may be the primary mechanism for reducing the number of LRDs within the observed redshift range. Cosmological \(N\)-body simulations find that major mergers with mass ratio \(\xi\gtrsim0.3\) occur with a frequency \(p\sim0.2\) per halo per unit redshift~\cite{Inayoshi:2025isg,Fakhouri:2010st}; the rate is nearly independent of redshift out to \(z\sim15\) and depends only weakly on halo mass. %(\(\propto M^{0.13}\)). 
The comoving number density of seed-induced haloes that never experience such mergers and thus retain their seed-induced features is
\(n_{\rm seed}(z)=n_{\rm seed,0}\exp\bigl[-p\,(z_0-z)\bigr]\),
where \(z_0\) is the initial redshift at which counting begins. 

As the abundance of \(\Lambda\)CDM (not PBHs) haloes is subdominant at \(z\gtrsim15\) compared to the PBH seed-induced haloes of similar size; and PBHs are also a tiny fraction $f_{\rm pbh}$ of dark matter, frequent mergers of seed-induced haloes are not expected for \(z\gtrsim15\). We may set \(p=0.2\) and \(z_0=15\), which results in \(\sim10\%\) of the seed-induced haloes surviving without major mergers until \(z\sim 4\textup{--}6\). Combining this survival fraction with the compactness selection above gives an expected comoving number density of observable LRDs, \(n_{\rm LRD}\sim10^{-5}\textup{--}10^{-4}{\rm\,cMpc}^{-3}\).

It is also possible that most SMBHs are dormant (small duty cycle $f_{\rm duty}$) and therefore do not exhibit AGN-like activity. If the observable fraction is set by \(f_{\rm duty}\sim0.01\textup{--}0.1\), then \(n_{\rm LRD}=f_{\rm duty}\,n_{\rm compact}\sim10^{-6}\textup{--}10^{-4}{\rm\,cMpc}^{-3}\) with \(n_{\rm compact}\sim10^{-4}\textup{--}10^{-3}{\rm\,cMpc}^{-3}\), similar to the above estimate. This may also explain why most SMBHs are not detected by JWST. However, the real situation is more complex and is likely driven by the combined effect of multiple mechanisms, including the two discussed above. 

Note that the above argument does not rely on the assumption that the abundance of faint galaxies traces the abundance of all dark haloes—an assumption \cite{bouwens2021new} that may break down at high redshift. Instead, our estimate follows directly from the PBH abundance together with the selection criteria for LRDs.

\subsection{Redshift evolution of LRDs}
The abundance of LRDs is found to decline rapidly around $z\sim4$. The decline may reflect both the selection effects and the evolution of LRD morphology~\cite{Ma:2025qna}. Physically, it can be understood if the LRD phase is a transient early black-hole growth stage: dense gas around a newly formed seed black hole can produce the red optical continuum, Balmer absorption/break features, X-ray weakness, and super-Eddington growth, but after one or two accretion episodes the gas supply is reduced or expelled and the source evolves into a more normal AGN~\cite{Inayoshi:2025isg}. In parallel, host-galaxy growth can hide the LRD signature: inside-out growth and cold accretion build blue star-forming outskirts around the compact red core, increasing the effective size and weakening the V-shaped spectral energy distribution used to select LRDs~\cite{kocevski2024rise,billand_investigating_2025}. Environment and halo evolution provide a related route: as LRD halos grow from \(\sim10^{10}{\rm\,M}_\odot\) at high redshift to \(\sim10^{11.3}{\rm\,M}_{\odot}\) by \(z \sim 3.5\), the systems become larger, less compact, more similar to ordinary galaxies, and their spectral energy distributions are altered by dense-gas depletion and enhanced star formation~\cite{zhang2026little}.

Recent studies find that there exist some LRD-like objects at lower redshifts. 
For instance, Ref.\,\cite{hviding2026xraydotexoticdust} reported an X-Ray Dot at \(z=3.28\) with LRD-like properties but unusually strong X-ray emission, consistent with an LRD in transition from an optically thick envelope to an AGN. Likewise, Ref.\,\cite{fu2025discovery} identified two LRDs at \(z=2.868\) and \(z=2.925\) that satisfy the LRD criteria yet already show enhanced X-ray, radio, and mid-IR emission, suggesting evolution toward typical AGNs. Similar LRD-like sources have also been found at lower redshifts \cite{Lin:2025pnq,ji2026lord,Chen:2025pgw,Rodriguez:2026zmc}. Together, these results indicate that LRDs indeed evolve into later AGN stages at lower redshifts, linking early SMBH seeds to local SMBHs.

\section{Conclusion} 
\label{Sec:conclusion}
We investigate the origin of the high-redshift SMBHs in LRDs observed by JWST from small-scale clustering of PBHs followed by runaway mergers. The derived LRD mass function in the small-scale PBH-clustering scenario agrees with the JWST observations for $1\textup{--}10\%$ accretion duty cycle. Other properties of LRDs—including their overmassive nature and compactness—are also explained within our framework when the seed effect and the halo spin distribution are taken into account. 

In addition, the dense gas residing in the PBH clusters is estimated to be consistent with the LRD observations of $\gtrsim10^{9}\text{--}10^{11}\, \mathrm{cm}^{-3}$. Moreover, the high-mass tail \(M_{\bullet}\gtrsim10^8{\rm\,M}_\odot\) of the SMBH population is crucial in accelerating early galaxy formation and can help explain the observed cumulative stellar mass density beyond the standard \(\Lambda\)CDM cosmology. Another distinctive signature of our model is the imprinted stochastic gravitational wave background during the PBH merger process in clusters~\cite{Zhang:2025tgm}, which can be distinguished from massive PBHs formed through direct collapse~\cite{Qin:2025ymc,Guo:2026cuv,Allegrini:2026jqt}; and other direct collapse channels associated with particle dark matter~\cite{Feng:2020kxv,Feng:2021rst,Jiang:2025jtr,Shen:2025evo,Feng:2025rzf,Feng:2025ybf,Aggarwal:2025pit,Gu:2026zzq,Bhattacharya:2025dgx}.

\bigskip

\section*{Acknowledgements}
This work is supported in part by the National Science Foundation of China (NSFC) under Grant Nos.\,12525506 and 12475107, the National Key R\&D Program of China under Grants Nos.\,2021YFC2203100 and 2017YFA0402204, the China Postdoctoral Science Foundation under Grant No.\,2024M761594, and the Shuimu Tsinghua Scholar Program.

\bibliographystyle{utphys}
\bibliography{LRDs}

\appendix

\end{document}